(Review Article)

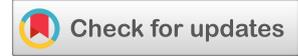

# Moving from monolithic to microservices architecture for multi-agent systems

Muskaan Goyal [1, *] and Pranav Bhasin [2]

[1] Department of Computer Science, University of California, Berkeley, US.
[2] Department of Electrical Engineering & Computer Science, University of California, Berkeley, US.

**Abstract**

The transition from monolithic to microservices architecture revolutionized software development by improving scalability and maintainability. This paradigm shift is now becoming relevant for complex multi-agent systems (MAS). This review article explores the evolution from monolithic architecture to microservices architecture in the specific context of MAS. It will highlight the limitations of traditional monolithic MAS and the benefits of adopting a microservices-based approach. The article further examines the core architectural principles and communication protocols, including Agent Communication Languages (ACLs), the Model Context Protocol (MCP), and the Application-to-Application (A2A) protocol. The article identifies emerging architectural patterns, design challenges, and considerations through a comparative lens of the paradigm shift.

**Keywords:** Microservices; Multi-Agent Systems; Multi-Agent Microservice Architecture; Model Context Protocol; Agent Communication Languages; Application-to-Application Protocol

## 1. Introduction

Monolithic architecture is a traditional software development approach where all application components, including user interface, business logic, and data access layer, are tightly integrated into a single, indivisible unit. As a monolithic approach is relatively more straightforward to set up, it is more common for smaller applications and teams during the early stages of software development. The inherent limitations of monolithic architecture have become apparent due to the ever-growing complexity of software systems [1, 2].

While monolithic architectures continue to offer simplicity and efficiency for many applications, the growing demands for scalability, modularity, and rapid iteration in complex systems have made microservices architecture an increasingly attractive alternative. This paradigm integrates an application in a decentralized approach where each independent service performs a distinct business function and communicates with others through well-defined interfaces and a network, often using lightweight protocols. This decomposition offers several key benefits, including enhanced scalability, improved maintainability, and the ability to deploy individual services independently [1, 2].

Just as the limitations of monolithic applications led to the rise of microservices to manage growing complexity and scale, a similar architectural shift is emerging in the era of large language models (LLMs). Instead of relying on a single general-purpose LLM to handle all tasks, systems are increasingly orchestrating groups of specialized agents. This approach enables them to address diverse goals and capabilities and has driven a renewed focus on multi-agent system (MAS) architectures [3].

With the advent of generative AI and large language models (LLMs), multi-agent systems (MAS) have become a prominent paradigm for developing complex systems. MAS usually consists of multiple autonomous agents that can interact with each other to achieve a common goal. Given the inherent decentralized nature of MAS, microservices architecture is a natural fit for their development and deployment [4].

[*] Corresponding author: Muskaan Goyal

The adoption of microservices for MAS is driven by factors such as rapid development cycles involving frequent and independent updates to individual agents, scalable deployment, and efficient resource management.

## 2. The Evolution from Monolithic to Microservices Architecture

Historically, monolithic architectures have been a fundamental architectural approach in software development during the early stages of development.

With a single codebase and a unified deployment unit, software systems benefited from simple, straightforward development and deployment processes. Companies like Netflix and Airbnb initially operated on monolithic systems because of easy testing and debugging within a centralized environment [5].

The limitations of the monolithic architecture became clear as applications grew in complexity and systems suffered from inefficient scalability. If a specific component experiences high demand, monolithic applications require scaling the entire application, resulting in inefficient resource utilization and increased costs [6].

Furthermore, development cycles for monolithic architectures tend to be longer, which hinders agility. Adopting new technologies is also complex and time-consuming with a centralized approach because it requires updating the entire application.

As the monolithic application grew in size, maintainability became a significant concern. Updating and modifying became risky due to the tight coupling between different components [5]. A bug fix could impact the entire system, causing unnecessary code maintenance challenges. Such systems require more refactoring and are more likely to introduce new errors during updates, causing deployment risks. Lastly, monolithic systems suffer from poor fault isolation, where a failure in one component can bring down the entire application [5].

The limitations mentioned above motivated the shift to microservices architecture. Due to its decentralized approach, microservices architecture helped the organization achieve improved scalability, agility, faster development, and resilience.

## 3. Microservices Architecture Principles Relevant to Multi-Agent Systems

Microservices are a natural fit for multi-agent systems (MAS) due to their fundamental characteristics.

The principle of decentralization in microservices resonates with the concept of autonomy in MAS. The microservice architecture allows each agent to operate as an independent entity capable of making decisions and taking actions without centralized control [7]. It will enable individual agents in MAS to be designed and implemented independently, which fosters flexibility and allows for concurrent development efforts [8].

The principle of Independent Deployment in microservices can be extended for MAS, where specific agents might require frequent updates with newer models or fine-tuned variations. It improves development agility by ensuring that updates can be rolled out efficiently without disrupting the entire application [1].

Similarly, the principle of Bounded Contexts aligns with the MAS, where each agent is typically designed to perform a distinct business function or task [9]. A microservices-based MAS can use this principle to improve the application's maintainability by organizing it so that each agent is responsible for a specific area of expertise.

## 4. Monolithic Multi-Agent Systems: Limitations and the Need for Change

A monolithic MAS within an application characterizes all agents' interconnectedness and functionalities. This approach offers initial simplicity for small-scale MAS but has significant challenges as the number of agents increases and the system grows in complexity.

Similar to traditional monolithic systems, monolithic MAS faces difficulty in achieving scalability. If only a subset of agents requires additional resources, the entire monolithic application must be scaled. This unnecessary scaling would lead to inefficient resource utilization and increased operational costs [5].

As the number of agents grows, coordination complexity is a significant challenge in monolithic MAS. Managing the interactions between multiple tightly coupled agents and ensuring coherent behavior toward common goals is difficult. Complex dependency chains can lead to cascading exceptions and system instability. These exceptions also present a hurdle to effectively troubleshooting issues [5].

Monolithic MAS also faces hurdles due to compromised maintainability and upgradability. Introducing changes, new agent types, and adopting new AI models or technology can be risky and time-consuming because modifications to one agent might have unintended consequences on other parts of the application.

Lastly, monolithic MAS suffers from poor fault isolation due to widespread disruptions. A critical fatal error in one agent can impact the functionality of all other agents and cause the entire multi-agent application to crash [5].

A shift towards a modular, independent, distributed architecture like microservices becomes essential to overcome these limitations in MAS. This architecture shift is paving the way for more robust and adaptable agent-based applications.

## 5. Microservices Architecture for Multi-Agent Systems: Communication Protocols

When adopting a microservices-based architecture for MAS, it is vital to recognize the parallels between agents in MAS and services in microservice-based systems. Similar to microservices, each agent acts as an independent, specialized unit that can interact and collaborate with other agents to achieve the system's overall objectives [10].

Communication plays a key role in both microservices and MAS. In microservices, different services typically talk to each other through clearly defined interfaces using protocols like RESTful APIs or gRPC [11]. Similarly, agents in a MAS need to communicate to coordinate their behavior and work together toward common goals. Although standard APIs can be used, microservices-based MAS can also use specialized protocols for semantic interactions between autonomous components.

Several protocols play a crucial role in facilitating interaction between agents in microservices-based MAS:

- **Agent Communication Languages (ACLs)** such as FIPA offer a formal structure for agents to exchange messages with built-in semantics [12]. These ACLs define message formats and include performatives like "request" and "inform" to explicitly convey the purpose of communication [13]. This communication protocol allows the agents to understand and respond appropriately based on the context.
- **The Model Context Protocol (MCP)** is gaining traction as a standardized interface enabling seamless interaction and integration for AI models, agents, tools, and data sources. MCP is also often described as the "USB-C port for AI applications" [14]. It adopts a client-server architecture: AI applications (Hosts) connect via Clients to various context providers (Servers). Furthermore, it facilitates the transfer of Tools (executable functions), Resources (data sources), and Prompts (predefined templates) to enrich the decision-making and response generation quality of AI models by injecting them with real-time, relevant external context [15].
- **The Application-to-Application (A2A) Protocol** is a critical communication layer between intelligent agents or applications, complementing MCP [16]. It emphasizes coordination and collaboration, introducing key concepts like the *Agent Card*, which advertises an agent's capabilities, and the *Task* object, which encapsulates discrete units of work shared between agents [17].

While communication protocols in traditional microservices revolve around the exchange of well-structured data payloads such as JSON formats, the protocols in MAS extend beyond raw data exchange, which involves transmitting explicit semantic intentions or goals. For example, rather than merely sending a search query, an agent might instruct another to search a database for articles, embedding the purpose of the action directly within the message. This kind of semantic communication fosters more intelligent collaboration because agents can interpret and act on the intent behind messages.

## 6. Architectural Patterns in Multi-Agent Systems

Beyond the fundamental principles of microservices and the core communication protocols like MCP and A2A, several architectural patterns are proving effective for designing and implementing microservices-based multi-agent systems.

- **Orchestrator-Worker Pattern**: This pattern features a central orchestrator agent responsible for delegating tasks to a pool of worker agents and overseeing their execution to achieve a shared objective [18]. It introduces a centralized control mechanism for task decomposition and coordination. The design can be enhanced through an event-driven approach using platforms like Kafka. The orchestrator emits tasks as events, and worker agents subscribe to and act on those events asynchronously [19].
- **Hierarchical Agent Pattern**: In this pattern, agents are arranged in a tiered structure, with higher-level agents supervising and delegating work to those below them. This hierarchy is particularly effective for tackling large-scale, complex problems by recursively breaking them into smaller, manageable components [20]. Each node (except the leaves) functions as an orchestrator for its sub-agents, and like the orchestrator-worker model, it can benefit from event-driven implementation strategies.
- **Multi-Agent MicroServices (MAMS) Architecture**: MAMS offers a dedicated architecture for integrating MAS within a microservices framework. It distinguishes between Agent-Oriented MicroServices (AOMS), which leverage MAS frameworks and house communities of autonomous agents, and Plain-Old MicroServices (POMS), which expose traditional RESTful APIs [21]. In this setup, MAMS agents serve as interface agents, collectively forming the microservice's REST layer with which external services can interact, effectively bridging MAS and conventional service ecosystems [10].

In addition to the above-established architectural patterns, research has highlighted several other design strategies, such as the Broker Pattern and Blackboard Pattern, from the broader MAS domain that can be effectively applied within a microservices context.

## 7. Comparative Analysis

It is crucial to compare microservices designed for non-agentic systems with those designed for agentic systems to adopt a microservices architecture for multi-agent systems efficiently.

**Table 1** Microservices for Non-Agentic Systems vs. Microservices for Agentic Systems

| Feature | Microservices for Non-Agentic Systems | Microservices for Agentic Systems |
| --- | --- | --- |
| Communication Paradigm | Services primarily use REST, GraphQL, or gRPC APIs for data exchange [11]. | Agents may use traditional APIs but often incorporate specialized protocols like ACLs, MCP, and A2A to facilitate semantic communication and context sharing [12, 15]. |
| State Management | Services are designed to be stateless, whereas external databases typically manage the state [21]. | Agents require maintaining internal state and context over time and across interactions. This requirement can be satisfied by incorporating specific strategies for distributed state management. |
| Orchestration | Service interactions are coordinated using centralized orchestration or event-based choreography. | Autonomous agents can involve both centralized and decentralized orchestration for coordination and negotiation [18]. |
| Autonomy | Services typically operate reactively—responding to incoming requests and executing predefined workflows based on those inputs. | Agents exhibit a degree of autonomy, allowing them to independently decide, plan, and act to achieve common goals. |
| Interaction Models | Interactions primarily consist of request-response based on clearly defined API contracts [22]. | Interactions extend beyond simple request-response patterns to complex, goal-driven exchanges, including negotiation and collaborative planning. |

## 8. Design Challenges and Considerations

There are several unique challenges to designing and implementing a microservices architecture for a multi-agent system.

- **State Management:** Historically, microservices were often designed to be stateless to improve scalability [23]. In a MAS, agents typically need to maintain internal state and context over time and across interactions, which presents a unique challenge for the microservices architecture [24]. The design must incorporate mechanisms to manage and persist the state information in a distributed environment.
- **Latency:** A microservices-based MAS application can suffer from an overhead latency due to frequent message passing between multiple agents and inter-service communication. Ensuring efficient performance, especially in real-time systems, hinges on streamlining communication protocols and minimizing network hops. These optimizations help reduce latency, prevent bottlenecks, and support timely, responsive interactions across distributed components [25].

## 9. Conclusion

The shift from monolithic to microservices architecture marks a key evolution in the design and deployment of software applications, and this evolution is now extending to multi-agent systems. Adopting microservices for multi-agent systems has numerous benefits, such as enhanced stability, improved maintainability, and increased agility. Specialized communication protocols like Model Context Protocol and Application-to-Application enable effective interaction and seamless collaboration among autonomous AI agents operating within a microservices architecture.

However, this approach faces several design challenges. Teams must carefully manage system state, address latency issues, ensure proper debugging capabilities, and maintain security. As generative AI advances, the demand for innovative, scalable multi-agent systems grows stronger. The microservices architecture is set to become a key framework for building such applications because it offers decentralization, clear communication, and modular design. These core microservices principles are being adapted and improved to handle the unique demands of generative AI applications.

## Compliance with ethical standards

*Disclosure of conflict of interest*

There are no conflicts of interest to be disclosed.

## References

[1] Newman S. Building microservices: designing fine-grained systems. " O'Reilly Media, Inc."; 2021 Jul 24.

[2] Dragoni N, Giallorenzo S, Lafuente AL, Mazzara M, Montesi F, Mustafin R, Safina L. Microservices: yesterday, today, and tomorrow. Present and ulterior software engineering. 2017:195-216.

[3] Dorri A, Kanhere SS, Jurdak R. Multi-agent systems: A survey. IEEE Access. 2018 Apr 18;6:28573–93. doi:10.1109/access.2018.2831228

[4] Wu Q, Bansal G, Zhang J, Wu Y, Li B, Zhu E, Jiang L, Zhang X, Zhang S, Liu J, Awadallah AH. Autogen: Enabling next-gen llm applications via multi-agent conversation. arXiv preprint arXiv:2308.08155. 2023 Aug 16.

[5] Newman S. Monolith to microservices: evolutionary patterns to transform your monolith. O'Reilly Media; 2019 Nov 14.

[6] Kamisetty A, Narsina D, Rodriguez M, Kothapalli S, Gummadi JC. Microservices vs. Monoliths: Comparative Analysis for Scalable Software Architecture Design. Engineering International. 2023;11(2):99-112.

[7] Carabelea C, Boissier O, Florea A. Autonomy in multi-agent systems: A classification attempt. In International Workshop on Computational Autonomy 2003 Jul 14 (pp. 103-113). Berlin, Heidelberg: Springer Berlin Heidelberg.

[8] Park S, Sugumaran V. Designing multi-agent systems: a framework and application. Expert Systems with Applications. 2005 Feb 1;28(2):259-71.

[9] Seedat M, Abbas Q, Ahmed N, Feroz I, Qureshi A, Amelio A. Transition Strategies from Monolithic to Microservices Architectures: A Domain-Driven Approach and Case Study. VAWKUM Transactions on Computer Sciences. 2024 Jun 3;12(1):94-110.

[10] W. Collier R, O'Neill E, Lillis D, O'Hare G. MAMS: Multi-Agent MicroServices. In Companion proceedings of the 2019 world wide web conference 2019 May 13 (pp. 655-662).

[11] Niswar M, Safruddin RA, Bustamin A, Aswad I. Performance evaluation of microservices communication with REST, GraphQL, and gRPC. International Journal of Electronics and Telecommunication. 2024 Jun 20;70(2):429-36.

[12] Pitt J, Mamdani A. Communication protocols in multi-agent systems: a development method and reference architecture. InIssues in agent communication 2000 Jan 1 (pp. 160-177). Berlin, Heidelberg: Springer Berlin Heidelberg.

[13] Kone MT, Shimazu A, Nakajima T. The state of the art in agent communication languages. Knowledge and Information Systems. 2000 Aug;2:259-84.

[14] Kone MT, Shimazu A, Nakajima T. The state of the art in agent communication languages. Knowledge and Information Systems. 2000 Aug;2:259-84.

[15] Hou X, Zhao Y, Wang S, Wang H. Model Context Protocol (MCP): Landscape, Security Threats, and Future Research Directions. arXiv preprint arXiv:2503.23278. 2025 Mar 30.

[16] Announcing the agent2agent protocol (A2A) [Internet]. [cited 2025 Apr 21]. Available from: https://developers.googleblog.com/en/a2a-a-new-era-of-agent-interoperability/

[17] Google. GitHub - google/A2A: An open protocol enabling communication and interoperability between opaque agentic applications. [Internet]. GitHub. 2025 [cited 2025 Apr 22]. Available from: https://github.com/google/A2A

[18] Fourney A, Bansal G, Mozannar H, Tan C, Salinas E, Niedtner F, Proebsting G, Bassman G, Gerrits J, Alber J, Chang P. Magentic-one: A generalist multi-agent system for solving complex tasks. arXiv preprint arXiv:2411.04468. 2024 Nov 7.

[19] Four Design Patterns for Event-Driven, Multi-Agent Systems [Internet]. Confluent. 2025 [cited 2025 Apr 22]. Available from: https://www.confluent.io/blog/event-driven-multi-agent-systems/

[20] Geng M, Pateria S, Subagdja B, Tan AH. HiSOMA: A hierarchical multi-agent model integrating self-organizing neural networks with multi-agent deep reinforcement learning. Expert Systems with Applications. 2024 Oct 15;252:124117.

[21] Laigner R, Zhou Y, Salles MA, Liu Y, Kalinowski M. Data management in microservices: State of the practice, challenges, and research directions. arXiv preprint arXiv:2103.00170. 2021 Feb 27.

[22] Chandramouli R. Microservices-based application systems. NIST Special Publication. 2019 Aug;800(204):800-204.

[23] Mooghala S. A Comprehensive Study of the Transition from Monolithic to Micro services-Based Software Architectures. Journal of Technology and Systems. 2023;5(2):27-40.

[24] Jagutis M, Russell S, Collier RW. Using multi-agent microservices (mams) for agent-based modelling. International Workshop on Engineering Multi-Agent Systems 2023 May 29 (pp. 85-92). Cham: Springer Nature Switzerland.

[25] Tanenbaum AS, Van Steen M. Distributed systems. CreateSpace Independent Publishing Platform; 2017 Feb.